# Moments of the Cluster Distribution as a Test of Dark Matter Models

**Manolis PLIONIS**[1], **Stefano BORGANI**[2,1]
**Lauro MOSCARDINI**[3] & **Peter COLES**[4]
[1] *SISSA - International School for Advanced Studies, via Beirut 2-4, I-34013 Trieste, Italy*
[2] *INFN – Sez. di Perugia, c/o Dip. di Fisica dell'Università, via A. Pascoli, I-06100 Perugia, Italy*
[3] *Dipartimento di Astronomia, Università di Padova, vicolo dell'Osservatorio 5, I-35122 Padova, Italy*
[4] *Astronomy Unit, School of Mathematical Sciences, Queen Mary & Westfield College, Mile End Road, London E1 4NS, UK*

# Abstract

We estimate the variance and the skewness of the smoothed density field of simulated clusters in several dark matter (DM) models. The cluster simulations are based on the Zel'dovich approximation, the low computational cost of which allows us to run as many as 50 random realizations of each model. The simulated cluster distributions are smoothed with a Gaussian window at two different smoothing scales, $R_{sm} = 20$ and $30\,h^{-1}$Mpc. We compare our results with a similar analysis of a redshift sample of Abell/ACO clusters (Plionis & Valdarnini 1994). Within the list of considered models, we find that only the mixed Cold+Hot DM model (with $\Omega_{\rm hot} = 0.3$) provides a good fit to the data. The standard CDM model and low–density ($\Omega_\circ = 0.2$) CDM models, both with and without a cosmological constant term ($\Omega_\Lambda = 0.8$), are ruled out. The tilted CDM model with primordial spectral index $n = 0.7$ and a low Hubble constant ($h = 0.3$) CDM model are only marginally consistent with the data.

# 1 Introduction

On scales traced by galaxy clusters ($\gtrsim 20\,h^{-1}$Mpc), gravitational clustering has not yet entered its non–linear phase of evolution, so that the cluster distribution is mainly determined by the shape of the primordial fluctuation spectrum[1]. For this reason, several attempts have been devoted in compiling extended cluster redshift surveys both in the optical (e.g., Postman, Huchra & Geller 1992; Dalton et al. 1994, and references therein) and in the X–ray (e.g., Nichol, Briel & Henry 1994) bands. In order to compare the observational data sets with different cosmological models several authors have resorted to large N–body simulations which were designed to sample the length-scales relevant to the cluster distribution (e.g., White et al. 1987; Bahcall & Cen 1992; Croft & Efstathiou 1994). Since, however, linear, or at most mildly non–linear, gravitational clustering governs the cluster distribution, an N–body based approach is probably more sophisticated than what is really needed. On the other hand, existing analytical approaches, based either on linear theory (e.g., Bardeen et al. 1986; Holtzman & Primack 1993) or on the Zel'dovich approximation (cf., Doroshkevich & Shandarin 1978; Mann, Heavens & Peacock 1993) rely on simplifying assumptions which do not allow an accurate treatment of the high–order correlation statistics or a realistic account to be taken of the observational biases, present in real data sets.

In this *Letter* we analyze cluster simulations based on the Zel'dovich approximation (Zel'dovich 1970). It has been shown that this kind of simulation gives a reliable description of the cluster distribution at a fraction of the computational cost of N–body simulations (Borgani, Coles & Moscardini 1994; Borgani et al. 1994b; see also Blumenthal, Dekel & Primack 1988). Taking advantage of this, we analyze six different initial power–spectra and, at the same time, run a large number (50) of realizations of each model to get a reliable estimate of the cosmic variance on the correlation statistics.

We analyze the density field traced by cluster distributions, after suitably smoothing it at different scales. This procedure has been recently applied to the cluster distribution by different authors (Plionis & Valdarnini 1994; Kolatt, Dekel & Primack 1994), since it limits the effect of shot–noise in clustering measures while a different approach, based on counts in cells and on measures of volume–averaged correlation functions, has been applied by Cappi & Maurogordato (1994). We evaluate the variance and the skewness of the resulting cluster density field. The comparison between our results and those obtained by Plionis & Valdarnini (1994; PV94 hereafter), using a similar analysis on an Abell/ACO redshift sample, allows us to put stringent constraints on the dark matter (DM) models we have considered. This is a further confirmation (see Borgani, Coles & Moscardini 1994) that our method to generate realistic cluster distributions represents a flexible instrument which can be used effectively to test the parameter space of DM models against the observed large–scale cluster distribution.

[1] $h$ is the Hubble constant $H_0$ in units of 100 km s$^{-1}$ Mpc$^{-1}$.

The layout of this *Letter* is as follows: in Section 2 we describe briefly our simulations and the considered models; in Section 3 we present the analysis method and the results obtained; we state our main conclusions in Section 4.

## 2 The Simulations

Here we simply sketch our simulation procedure, based on the Zel'dovich approximation [ZA] (cf. Shandarin & Zel'dovich 1989); more details are given in Borgani et al. (1994b).

Let $\mathbf{q}$ and $\mathbf{x}(\mathbf{q},t)$ be the initial (Lagrangian) and the final (Eulerian) comoving position of a fluid element. The ZA assumes that, at the time $t$, the Eulerian position is given by $\mathbf{x}(\mathbf{q},t) = \left[\mathbf{q} + b(t)\ \nabla_{\mathbf{q}}\psi(\mathbf{q})\right]$. Here $b(t)$ is the growing mode of the evolution of linear density perturbations and $\psi(\mathbf{q})$ is the gravitational potential, which is related to the initial (linear) density fluctuation field, $\delta(\mathbf{q})$, through the Poisson equation. Accordingly, the fluid particles move under the ZA along straight lines. Although gravity determines the initial "kick" given to the fluid particles, they do not afterwards feel any tidal effects. Particles fall into gravitational potential wells to form structures, but these structures evaporate subsequently since no account is taken of the self–gravity of these non–linear structures. In reality particles will attract each other when their orbits cross or they pass close to each other but, in the ZA, matter always moves along its initial path, oblivious to the presence of the other particles. In this sense, the ZA gives a good description of gravitational dynamics as far as particle trajectories do not intersect with each other, i.e. before "shell–crossing": after this moment, the validity of the ZA breaks down.

In order to improve the performance of ZA, one can filter out the small–scale wavelength modes in the linear power–spectrum, $P(k)$, which are responsible for shell–crossing (Coles, Melott & Shandarin 1993; see also Kofman et al. 1992). Accordingly, we convolve the linear power–spectrum with a Gaussian filter, $P(k) \to P(k)\, e^{-(kR_f)^2}$, which has been shown (Melott, Pellman & Shandarin 1994) to be the optimal choice. Kofman et al. (1994) provided an analytical expression for the average number of streams, $N_s$, at each Eulerian point as a function of the r.m.s. fluctuation amplitude, $\sigma$, of the Gaussian density field. The limit corresponding to the single–stream regime, $N_s \to 1$, is attained for $\sigma \to 0$, with $N_s$ increasing rapidly at larger values of $\sigma$: see eq.(27) in Kofman et al. (1994). We choose the filtering radius $R_f$ for each model so that $N_s = 1.1$. We found this choice to be a reasonable compromise between avoiding shell–crossing and suppressing the development of genuine clustering. In the second column of Table 1 we report the filtering radii used for the DM models considered in this paper.

We have generated random–phase realizations of the linear density field on $128^3$ grid points for a box of side $L = 320\,h^{-1}$Mpc. We moved particles from initial grid positions according to the ZA and re-assigned the density and the velocity field on the grid through a TSC interpolation scheme (see e.g. Hockney & Eastwood 1981) for the mass and the momentum carried by each particle. Then we defined clusters as local density maxima on the grid according to the following prescription. If $d_{cl}$ is the average cluster separation, then we select $N_{cl} = (L/d_{cl})^3$ clusters at the $N_{cl}$ highest density peaks. In the following, we assume $d_{cl} = 40\,h^{-1}$Mpc, which is the corresponding value of the combined Abell/ACO cluster sample to which we will compare our simulation results (see PV94).

We ran simulations for six different models of the initial fluctuation spectrum. For each model we generated 50 random realizations, in order to estimate the cosmic variance reliably. The models we have considered are the following: the standard CDM model (SCDM) with $\sigma_8 = 1$ for the r.m.s. fluctuation amplitude within a top–hat sphere of $8\,h^{-1}$Mpc; a tilted CDM model (TCDM), with primordial spectral index $n = 0.7$, $h = 0.5$ and $\sigma_8 = 0.5$ (e.g., Cen et al. 1992; Tormen et al. 1993; Adams et al. 1993; Moscardini et al. 1994); a low Hubble constant CDM model (LOWH), with $h = 0.3$ and $\sigma_8 = (1.6)^{-1}$ (Bartlett et al. 1994); a Cold+Hot DM model (CHDM), with $\Omega_{\rm hot} = 0.3$ for the fractional density of the hot component, $h = 0.5$ and $\sigma_8 = (1.5)^{-1}$ (e.g., Klypin et al. 1993); an open CDM model (OCDM), with $\Omega_\circ = 0.2$, $h = 1$ and $\sigma_8 = 1$; a spatially flat, low–density CDM model ($\Lambda$CDM), with $\Omega_\circ = 0.2$, $h = 1$, $\sigma_8 = 1.3$ and cosmological constant $\Omega_\Lambda = 0.8$ (Bahcall & Cen 1992; Dalton et al. 1994). Transfer functions have been taken from Holtzman (1989), except that of LOWH, which was taken from Bond & Efstathiou (1984), with suitably chosen shape parameter $\Gamma = \Omega_\circ h = 0.3$. All the models, except the open CDM one, are normalized to be consistent with the quadrupole of the CMB temperature anisotropy measured by COBE (Bennett et al. 1994). In a forthcoming paper (Borgani et al. 1994b), we will present more details about our simulations.

# 3 The analysis

In the following, we implement the same analysis as in PV94, in order to allow a reliable comparison between real cluster data and our simulations (we refer to that paper for further details on the the real cluster sample and the analysis method; see also Borgani et al. 1994b). We obtain a continuous cluster density field by smoothing the cluster distribution on a grid, with grid-cell width of 20 $h^{-1}$ Mpc, using a Gaussian kernel. In order to study the cluster density field at different smoothing scales, $R_{sm}$, we use two radii for the Gaussian kernel, namely $R_{sm} = 20$ and 30 $h^{-1}$ Mpc. If $\rho(\mathbf{x}_g, R_{sm})$ is the smoothed density field at the grid point position $\mathbf{x}_g$, then the relative fluctuations are given by $\delta(\mathbf{x}_g, R_{sm}) = \rho(\mathbf{x}_g, R_{sm})/\bar{\rho} - 1$, where the average density $\bar{\rho}$ does not depend on $R_{sm}$. The variance $\sigma^2(R_{sm})$ and the skewness $\gamma(R_{sm})$ are defined as the second– and third–order moments of the $\delta$ field, respectively. We

therefore have

$$\sigma^2(R_g) \;=\; \frac{1}{N_{gr}}\sum_{g=1}^{N_{gr}}\left[\delta(\mathbf{x}_g, R_{sm})\right]^2 \qquad ; \qquad \gamma(R_g) \;=\; \frac{1}{N_{gr}}\sum_{g=1}^{N_{gr}}\left[\delta(\mathbf{x}_g, R_{sm})\right]^3 \;, \qquad (1)$$

where $N_{gr} = (16)^3$ is the total number of grid points. We have chosen not to take into account Poisson shot–noise corrections since the cluster distribution can be hardly considered a Poisson sampling of the underlying (galaxy) distribution (cf. Coles & Frenk 1991; Borgani et al. 1994a). Moreover, Gaztañaga & Yokoyama (1993) have shown that the smoothing process itself suppresses considerably the shot–noise effects. For these reasons PV94 did not use any shot–noise corrections so, to make a consistent comparison of our models with their results, we did not include such corrections in our analysis either. Furthermore, since all the model cluster distributions have the same $\langle\rho\rangle$ and we treat them similarly, the possible effects of shot–noise are *relatively* and *qualitatively* canceled out in the model and data intercomparison.

In Figure 1 we plot the results of our analysis in the $\sigma^2$–$\gamma$ plane, following Coles & Frenk (1991; see also Coles et al. 1993). In each panel, we show the variance–skewness relation estimated in redshift space and at the two smoothing scales for the corresponding model. The analysis has been done in redshift space in order to compare our results properly with real data. Redshift distortions affect the $\sigma^2$ and $\gamma$ values by $\sim 10\%$ at $R_{sm} = 20\,h^{-1}$Mpc and by $\sim 6\%$ at $R_{sm} = 30\,h^{-1}$Mpc (see also Borgani et al. 1994b). Each point represents one of the 50 realizations, so that the scatter represents the effect of cosmic variance. The crosses represent the results of the PV94 analysis of the combined Abell/ACO cluster redshift sample. In Table 1 we report, for each model, the values of $\sigma^2$ and $\gamma$ estimated as the average between the 50 realizations and their "cosmic r.m.s." uncertainties; we also give the corresponding PV94 results.

From Figure 1 we see that $R_{sm} = 20\,h^{-1}$Mpc is the most effective of the two scales we have considered for discriminating between the models. At this scale only the CHDM model fits best the data. The LOWH and TCDM are marginally consistent, with only about one tenth of the realizations giving $\sigma^2$ and $\gamma$ values as high as those of the Abell/ACO sample. The SCDM, $\Lambda$CDM and OCDM are ruled out, giving rise to either a too weak (SCDM) or a too strong clustering (OCDM and $\Lambda$CDM). The results at $R_{sm} = 30\,h^{-1}$Mpc are less discriminative. At this scale, the only models which can be ruled out to a high confidence level are the $\Lambda$CDM and OCDM for which, again, no realization reproduce the PV94 result. Although the other models are all consistent with observations, CHDM and TCDM seem to fare better (see also Table 1). We can conclude, therefore, that $R_{sm} = 20\,h^{-1}$Mpc is a sort of optimal scale where to test dark matter models against the distribution of galaxy clusters. At smaller scales shot–noise dominates in the estimation of the cell–count moments, while at larger scales the clustering becomes significantly weaker, and consequently more difficult to measure from a finite sample, be it real or simulated.

# 4 Discussion & Conclusions

We have used extensive simulations of rich galaxy clusters, based on the Zel'dovich approximation. We considered six different models for the linear power–spectrum and generated a large number (50) of random–phase realizations of each model, in order to determine accurately the effect of cosmic variance. We then generated a smoothed density field on a grid by convolving the cluster distributions with a Gaussian window at two different scales, $R_{sm} = 20$ and $30\,h^{-1}$Mpc. In order to constrain the power–spectrum models, we computed the variance and the skewness of the density field and compared it with the results of a similar analysis of an Abell/ACO combined cluster redshift sample, realised by Plionis & Valdarnini (1994).

We find that $R_{sm} = 20\,h^{-1}$Mpc is the most reliable scale at which to discriminate between models. At this scale, only the CHDM model fits the data very well. SCDM, $\Lambda$CDM and OCDM are ruled out at a high significance level. The LOWH and TCDM models are marginally consistent with the data. At $R_{sm} = 30\,h^{-1}$Mpc the only models to be definitely ruled out are $\Lambda$CDM and OCDM. Since the OCDM and $\Lambda$CDM have a very similar $P(k)$ shape, the higher normalization of the latter is the reason for the even stronger clustering it produces with respect to the former.

Clearly, one can imagine making modifications to the choice of the parameter values, involved in the definition of the DM models we have considered, in such a way as to make them consistent with the data. The physical justification for such an *ad hoc* tuning of parameters is, however, quite feeble at the present time. For example, in the LOWH model, it is quite hard to justify at the present time a Hubble parameter as small as $h = 0.3$ (Fukugita, Hogan & Peebles 1993). As far as TCDM is concerned, lowering the primordial spectral index still further, below $n \simeq 0.7$, could increase the excess power on large scales. However, a value even as low as $n = 0.7$ is strongly constrained by the level of CMB anisotropy detected by COBE (Bennett et al. 1994) and by the properties of the large–scale galaxy peculiar velocity field (e.g., Moscardini et al. 1994). Increasing the density parameter in the OCDM and $\Lambda$CDM models should decrease the cluster correlation to a satisfactory level: we have, in fact, verified that simulations based on $\Omega_\circ = 0.4$ fare much better than those we have considered here. In a forthcoming paper, we plan to exploit appropriate properties of the cluster velocity field in order to discriminate even more between high–density and low–density universes.

As a final comment we should mention that, although the CHDM model with $\Omega_{\rm hot} = 0.3$ is consistent with cluster correlation data, it has nevertheless been shown to be severely constrained by the detection of high–redshift objects (e.g. Klypin et al. 1994; Ma & Bertschinger 1994). Although lowering the hot fraction to $\Omega_{\rm hot} = 0.2$–$0.25$ may alleviate problems connected with the timing of galaxy formation in this scenario, it remains to be seen whether this can be achieved without destroying the apparent consistency of the model with the observed clustering of rich clusters that we have found in this work.

In the light of these further possible investigations, it is clear that the ease and low computational cost of our method for generating cluster simulations, provides an extremely useful tool for investigating the large parameter space of dark matter models and comparing them with real data.

## Acknowledgements

The authors wish to thank YiPeng Jing for pointing out an error in the normalization of $\Lambda$CDM in a previous version of this *Letter*. MP acknowledges the receipt of an EC *Human Capital and Mobility* Fellowship. PC acknowledges the receipt of a PPARC Advanced Research Fellowship. LM has been partially supported by Italian MURST. We are also grateful to PPARC for support under the QMW Visitors Programme in Astronomy GR/J 88357.

Table 1: Column 2: Filtering radius (in units of $h^{-1}$Mpc for the linear power–spectrum. Column 4 and 5: Variance and skewness in redshift space, at the two smoothing radii indicated in column 3 (in units of $h^{-1}$Mpc), for the six simulated models and for the real data (from PV94).

| Model | $\mathbf{R_f}$ | $\mathbf{R_{sm}}$ | $\sigma^2$ | $\gamma$ |
|---|---|---|---|---|
| **SCDM** | 4.4 | 20 | 0.373±0.029 | 0.262±0.060 |
| | | 30 | 0.110±0.014 | 0.023±0.011 |
| **TCDM** | 1.6 | 20 | 0.385±0.044 | 0.297±0.097 |
| | | 30 | 0.123±0.024 | 0.032±0.018 |
| **LOWH** | 2.4 | 20 | 0.385±0.042 | 0.282±0.080 |
| | | 30 | 0.119±0.020 | 0.027±0.014 |
| **CHDM** | 2.2 | 20 | 0.504±0.050 | 0.499±0.146 |
| | | 30 | 0.155±0.024 | 0.050±0.027 |
| **OCDM** | 4.5 | 20 | 0.638±0.058 | 0.771±0.189 |
| | | 30 | 0.215±0.031 | 0.089±0.039 |
| **ΛCDM** | 6.3 | 20 | 0.899±0.095 | 1.689±0.689 |
| | | 30 | 0.298±0.057 | 0.211±0.168 |
| **Abell/ACO** | | 20 | 0.496±0.064 | 0.445±0.187 |
| **(PV94)** | | 30 | 0.135±0.028 | 0.033±0.028 |

# Figure Caption

**Figure 1.** Scatter–plots for the variance–skewness relation. Each point refers to a single realization of the corresponding model. Triangles and squares are for $R_{sm} = 20\,h^{-1}$Mpc and $R_{sm} = 30\,h^{-1}$Mpc, respectively. Heavy crosses are the results of a similar analysis realized by PV94 on a combined Abell/ACO cluster redshift sample.

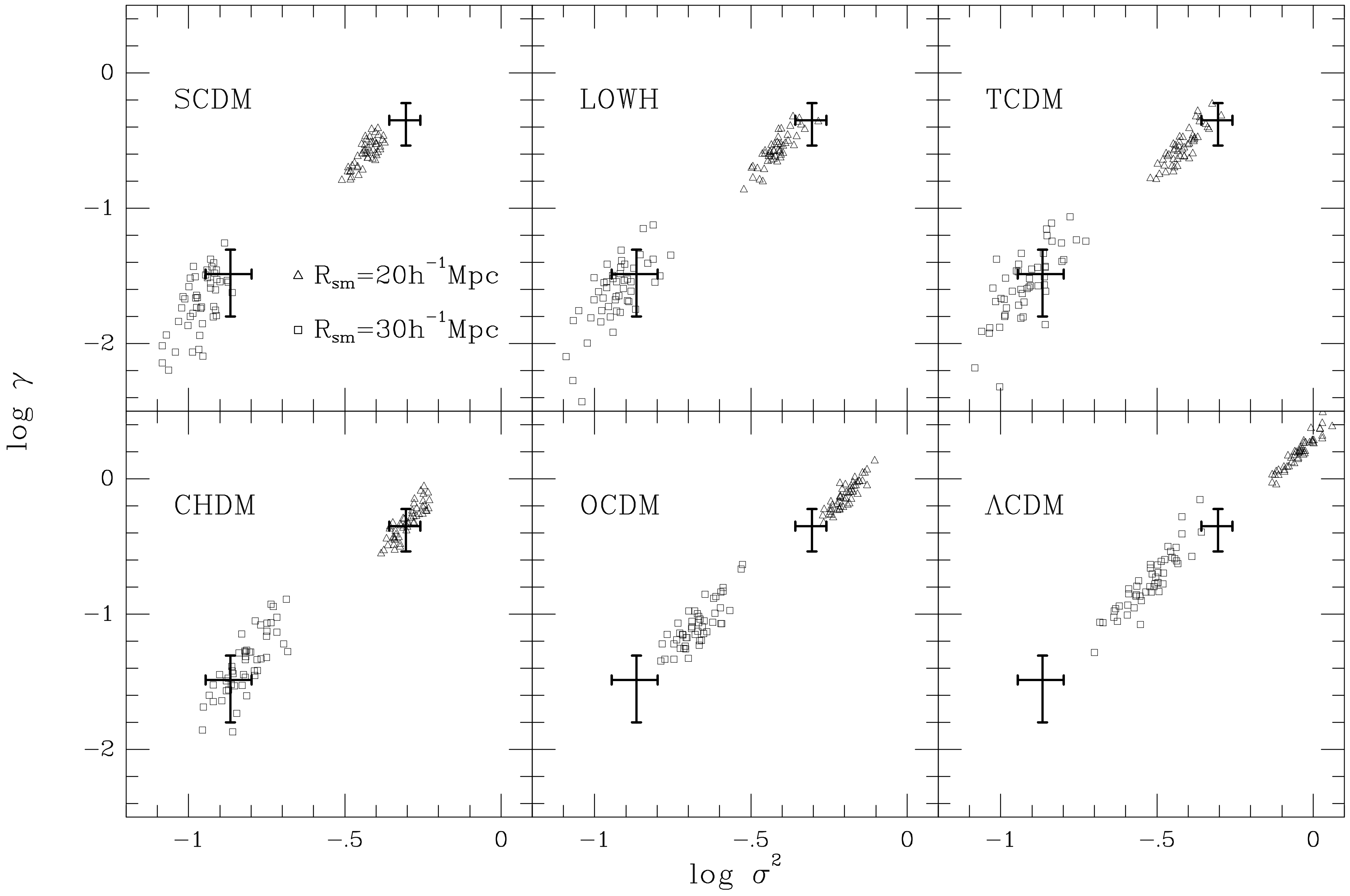